# Effect of high-κ environment on charge carrier mobility in graphene

L. A. Ponomarenko[1], R. Yang[1], T. M. Mohiuddin[1], M. I. Katsnelson[2], K. S. Novoselov[1], S. V. Morozov[1,3], A. A. Zhukov[1], F. Schedin[1], E. W. Hill[1], A. K. Geim[1]

[1]Manchester Centre for Mesoscience and Nanotechnology, University of Manchester, Oxford Road, Manchester M13 9PL, United Kingdom
[2]Institute for Molecules and Materials, Radboud University of Nijmegen, Toernooiveld 1, 6525 AJ Nijmegen, The Netherlands
[3]Institute for Microelectronics Technology, 142432 Chernogolovka, Russia

*It is widely assumed that the dominant source of scattering in graphene is charged impurities in a substrate. We have tested this conjecture by studying graphene placed on various substrates and in high κ media. Unexpectedly, we have found no significant changes in carrier mobility either for different substrates or by using glycerol, ethanol and water as a top dielectric layer. This suggests that Coulomb impurities are not the scattering mechanism that limits the mean free path attainable for graphene on a substrate.*

Graphene continues to attract massive interest, especially as a conceptually new electronic system [1]. One of the first but still unanswered questions about the electronic properties of graphene has been the question about the dominant source of scattering. What kind of impurities limits its mobility $\mu$ to typical values of ~10,000 cm$^2$/Vs which are currently achievable for graphene deposited on a substrate? The limited mobility severely hinders search for new phenomena and device applications and, without knowing the source of scattering, it is difficult to develop strategies for improving graphene's quality.

Immediately after the observation of the field effect in graphene [2], it was pointed out that the linear changes in its conductivity $\sigma$ as a function of gate voltage $V_g$ or carrier concentration $n$ could not be understood within the standard $\tau$ approximation because of the linear density of states in single-layer graphene [3]. On the other hand, the linear $n$ dependence could naturally be explained by charged impurities [3,4], which seemed an obvious candidate for being dominant scatterers in the one-atom-thick system unprotected from immediate environment and prone to chemical doping [5]. This conjecture agrees with the experiment on additional doping of graphene with potassium, in which $\mu$ decreased in the manner prescribed by theory, [6] and the recent measurements that show a drastic increase in $\mu$ for suspended samples [7]. Moreover, there is broad consensus that electron and hole puddles at the neutrality point (NP) [8] are caused by a background electrostatic potential and, therefore, it is tempting to attribute the puddles and the dominant scatterers to the same origin [4]. Still, there have been some unsettling observations that do not allow this straightforward explanation to become universally accepted. Among them is the fact that thermal annealing allows large shifts of the NP which are often not accompanied by any significant changes in $\mu$ (unless the initial mobility is low) [5]. Such behavior has been observed by many groups. Furthermore, single- and bi- layer graphene exhibit similar values of $\mu$ [1], whereas Coulomb impurities scatter differently in these materials (because of their different density of states), and the same impurity concentration should result in different $\mu$, the unsettling fact that remains unexplained.

In this Letter, we address the problem of dominant scatterers in graphene by employing two approaches. First, we point out that a common denominator in transport experiments on cleaved graphene has been the use of oxidized Si wafers, and the removal of the substrate has led to much higher $\mu$ [7]. This seems to suggest that impurities are in silicon oxide. To this end, we have studied devices placed on a number of different substrates but found the same typical $\mu$ as for graphene on SiO$_2$. Second, the strength of scattering by charged impurities should strongly depend on dielectric environment [3,4]. If ionized impurities were the limiting scatterers in today's standard devices with $\mu$ >5,000 cm$^2$/Vs, by covering them with glycerol

(dielectric constant $\kappa \approx 45$), ethanol ($\approx 25$) or water ($\approx 80$), $\mu$ should have increased by at least an order of magnitude, reaching above 100,000 cm$^2$/Vs. However, we observed only small changes in $\mu$ (typically, less than 20%), which shows that charged impurities are not the primary source of scattering.

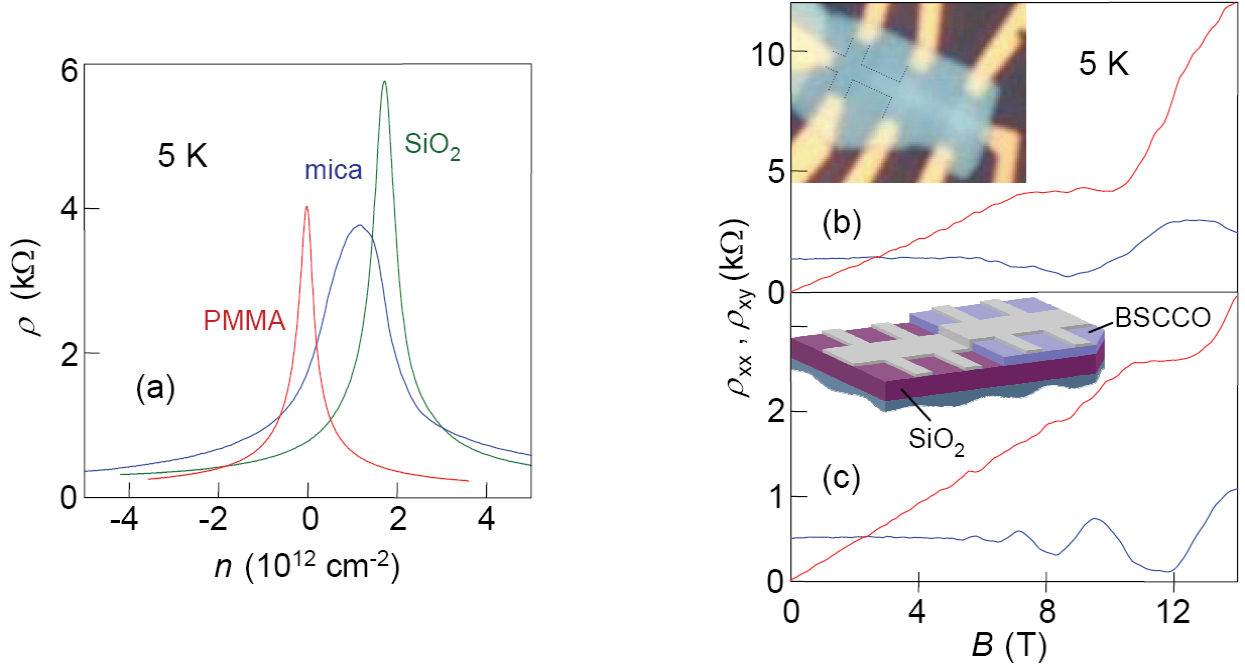

FIG. 1. Effect of the substrate on electron transport in graphene. (a) – Electric field effect for graphene on top of SiO$_2$, mica and PMMA. In each case, the induced carrier concentrations $n$ were related to $V_g$ through the gate capacitance [2] found by using Hall measurements. Positive and negative $n$ correspond to electrons and holes, respectively. (b,c) – $\rho_{xx}$ and $\rho_{xy}$ for graphene on BSCCO and SiO$_2$, respectively. These measurements were done at zero $V_g$ for the device made as shown schematically in (c): graphene extends over the edge of a thin BSCCO crystal to cover SiO$_2$. Upper inset: Optical micrograph of a graphene-on-BSCCO device. The bluish area is a BSCCO crystal on top of an oxidized Si wafer. Graphene is etched in the Hall bar geometry as indicated by the black lines on the left (in order to see the weak contrast due to graphene, we do not show such lines on the right). The width of the Hall bar is 1 μm.

The devices reported here were prepared from single-layer graphene deposited on various surfaces including SiO$_2$, polymethylmethacrylate (PMMA), spin-on glass (SOG), bismuth strontium calcium copper oxide (BSCCO), mica and boron nitride. In the case of SiO$_2$, SOG and PMMA, we used the standard procedure [2,5] to fabricate micron-sized Hall bar devices. PMMA and SOG were 100 nm thick, spun on top of an oxidized Si wafer (200 nm SiO$_2$) and cross-linked [9]. In the case of the other materials, we first prepared their crystallites (10 to 50 nm thick) on top of an oxidized Si wafer (300 nm SiO$_2$) by mechanical cleavage and then deposited graphene further on top (insets of Fig. 1). Fig. 1a plots examples of the electric field effect observed in graphene on different substrates (for each new material, at least 3 devices were measured). At high $n$, the shown resistivities $\rho$ vary by a factor of 4, yielding the same difference in $\mu$. To be specific, for typical $n \approx 2 \times 10^{12}$ cm$^{-2}$, the field-effect mobility $\mu_{FE} = \sigma/ne$ is $\approx$ 0.25, 0.45 and 0.8 m$^2$/Vs for the mica, SiO$_2$ and PMMA devices in Fig. 1a, respectively. Alternatively, we could use the expression $\rho = 1/ne\mu_L + \rho_S$ to discriminate between long and short range scattering that are described by constant $\mu_L$ and excess resistivity $\rho_S$ (both independent of $n$) [10]. The latter analysis yields $\mu_L \approx$ 0.25, 0.6 and 1.1 m$^2$/Vs and $\rho_S \approx$ 40, 110 and 60 Ω for mica, SiO$_2$ and PMMA, respectively. The variations in transport characteristics shown in Fig. 1a are well within the sample-to-sample variations typically observed for graphene-on-SiO$_2$ devices [1,11]. The only reproducible difference between the different substrates was that, for graphene on PMMA and SOG, the NP was always close to zero $V_g$ (without annealing) and the peaks were sharper than for the other substrates. It was generally possible to shift the NP to zero $V_g$ by in-situ annealing in He at 450 K, which as a rule did not result in notable changes unless the initial $\mu$ was low (<0.3 m$^2$/Vs).

An alternative way to assess the electronic quality was to apply magnetic field $B$ and measure the Hall mobility $\mu_H = \rho_{xy}/\rho_{xx}B$ where $\rho_{xx}$ and $\rho_{xy}$ are the longitudinal and Hall resistivities, respectively. Away from the NP, $\mu_H = \mu_{FE}$ according to both theory and experiment (see further). This approach was particularly suitable for graphene on BSCCO, which exhibited large hysteresis as a function of $V_g$. Figs. 1b,c show $\rho_{xx}$ and $\rho_{xy}$ which were measured for a device with graphene on both SiO$_2$ and BSCCO (inset of Fig. 1c). $\mu_H$ is given by the field at which $\rho_{xy} = \rho_{xx}$, yielding $\mu_H \approx 0.4$ m$^2$/Vs for both substrates.

Our efforts briefly described above show that the limited $\mu$ reported for graphene on SiO$_2$ are not due to charged impurities in the substrate as often speculated. However, the observed indifference with respect to substrates does not rule out charged impurities as the dominant source of scattering, because the experiments used similar microfabrication procedures and, in principle, one can imagine the same concentration of charged impurities always trapped underneath graphene [12]. To address the latter possibility, we have studied the influence of dielectric screening [13] on graphene's $\mu$.

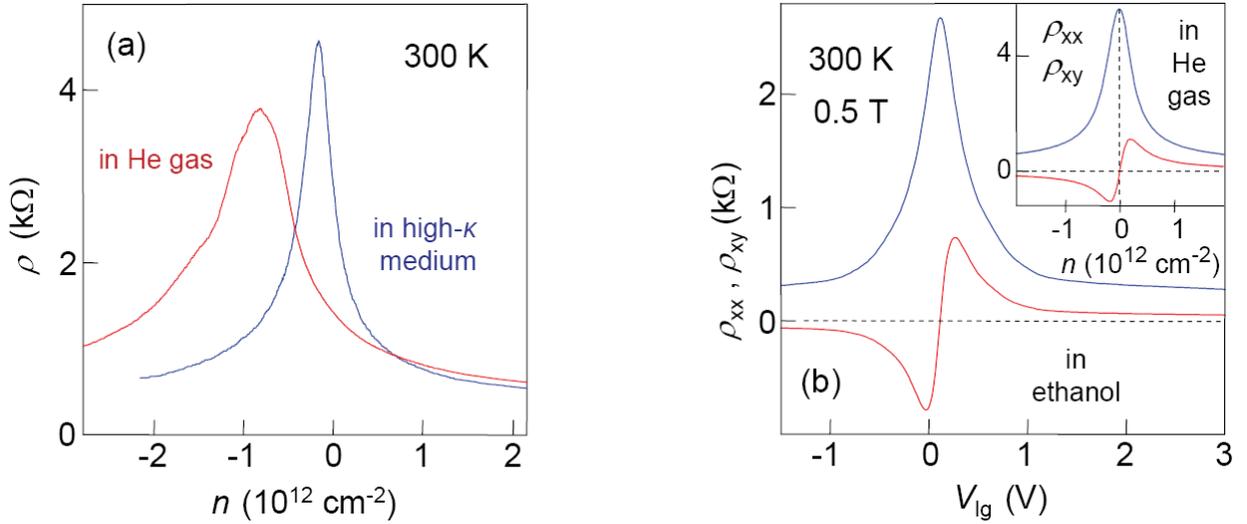

FIG. 2. Dielectric screening by high $\kappa$ media. (a) – Graphene in a helium atmosphere (red curve) and after covering it with a layer of glycerol at room $T$ (blue). The significant changes in the peak position and its width are *not* accompanied by the ten-fold increase in $\mu$, which would be expected for Coulomb scatterers [4,15]. The observed changes were even smaller for higher-$\mu$ devices. (b) – Another graphene device in ethanol and helium (inset) at room $T$. Simultaneous measurements of $\rho_{xx}$ and $\rho_{xy}$ (blue and red curves, respectively) allow us to find $n = B/\rho_{xy}e$ and $\mu_H = \rho_{xy}/B\rho_{xx}$ (see Fig. 3). Note that the experiments in ethanol were done by using the liquid gate so that one can imagine additional scatterers induced in the double layer by gate voltage. First, it is rather unlikely that the number of these scatterers exactly compensates the effect of dielectric screening and does this for all $n$. Second, the experiments using glycerol and the standard gating make this explanation even less likely. Nevertheless, let us mention that this doping effect can exist and was observed when we used a polymer electrolyte containing LiClO$_4$. In this case, $\mu$ did not stay constant but reduced very rapidly as $\propto 1/n$.

Fig. 2a plots the field effect in graphene on top of SiO$_2$, which was measured first in He atmosphere ($\kappa \approx 1$), then covered with a small droplet of glycerol ($\kappa \approx 42$ in the liquid state at room temperature $T$ and $\approx 65$ in the solid state at 220K) and measured again. The measurements were done in the standard geometry using the back gate [14]. In Fig. 2a we have intentionally chosen to show one of our lowest quality devices ($\mu_L \approx 0.5$ and 0.35 m$^2$/Vs for electrons and holes, respectively; $\rho_S \approx 80\pm20$ $\Omega$), in which the influence of $\kappa$ is clearly visible. One can see that glycerol significantly improved the device's characteristics making the peak in $\rho$ narrower, more symmetric and shifting it towards zero $V_g$. The features are consistent with the dielectric screening of charged impurities and electron-hole puddles. On the other hand, we have not observed the expected increase in $\mu$ by a factor of 10, which should have confirmed that charged impurities are the limiting scatterers. Note that the above factor takes into account the additional screening by graphene's charge carriers [4,15] and, without this screening, even greater enhancements in $\mu$ would be expected. For the device covered with glycerol, the curve in Fig. 2a yields $\rho_S \approx 100$ Ω and $\mu_L$ increases to $\approx 0.8$ (0.7) m$^2$/Vs for electrons (holes). This increase in $\mu$ by a factor of 1.6 (2) was the largest we have ever observed. For higher quality devices, typical increases in $\mu$ due to glycerol did not exceed 30%.

To corroborate these observations, we also used ethanol ($\kappa \approx 25$ at 300 K). In this case, it was difficult to use the back gate because of leakage currents, and we have found it more convenient to apply voltage directly to the liquid so that liquid-gate voltage $V_{lg}$ falls across a high-capacitance double layer at the graphene-ethanol interface, similar to the technique used for carbon nanotubes [16]. The capacitance of the double layer is a function of $T$, $V_{lg}$ and the amount of water and other impurities dissolved in ethanol and, therefore, in order to translate $V_{lg}$ into $n$, we simultaneously measured $\rho_{xy}$. Away from the NP ($n \geq 10^{12}$cm$^{-2}$), $\rho_{xy} = B/ne$ and data such as shown in Fig. 2b allow us to find both $n$ and Hall mobility $\mu_H$. We have first verified the equivalence of $\mu_H$ and $\mu_{FE}$ in a He atmosphere and found good agreement between the two types of measurements (Fig. 3a). For graphene in ethanol, one can see that $\mu_H$ goes higher and in parallel with respect to the measurements in He, yielding an increase in $\mu_L$ from $\approx 0.8$ to 0.9 m$^2$/Vs and little change in $\rho_S \approx 190 \pm 20$ Ω. We carried out such experiments for several devices and always observed higher $\mu$ in ethanol, with increases between few and 50% depending on graphene's initial quality. The higher the quality, the lower were the typical increases, in agreement with our observations for glycerol.

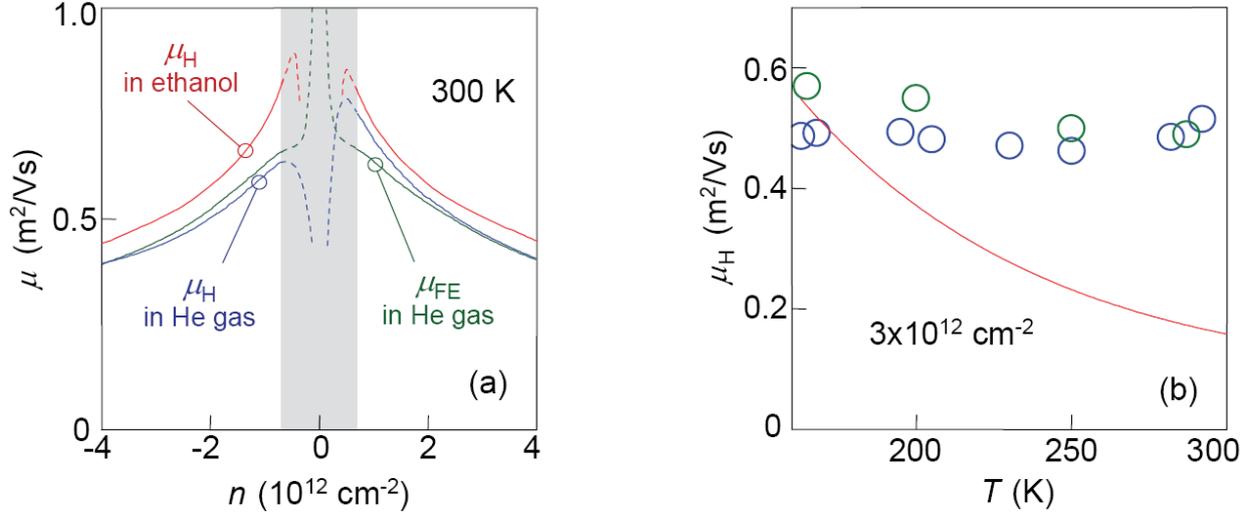

FIG. 3. Changes in $\mu$ induced by ethanol. (a) – Behavior of $\mu_H$ and $\mu_{FE}$ for the device shown in Fig. 2b. Measurements of $\mu$ are generally reliable only if $\mu_H \approx \mu_{FE}$. Inside the region around the NP, which is indicated by the gray area, apparent $\mu_{FE}$ diverges because $V_g$ passes through zero whereas apparent $\mu_H$ goes to zero because $\rho_{xy}$ passes through zero. The small difference between $\mu_H$ and $\mu_{FE}$ observed outside the gray area is attributed to a macroscopic inhomogeneity that leads to slightly different $\rho_{xy}$ for different pairs of Hall contacts. (b) – Varying dielectric screening in situ. $\mu$ as a function of $T$ for two devices immersed in ethanol (symbols). Dielectric constant $\kappa$ of ethanol increases from $\approx 25$ to 55 with decreasing $T$. The solid curve is the $T$ dependence expected in the case of dominant Coulomb scatterers [4,15]. The measurements are presented for $n = 3 \times 10^{12}$ cm$^{-2}$ but there is little difference for other assessable $n > 1 \times 10^{12}$ cm$^{-2}$. For the sample that shows a slight increase in $\mu_H$ at lower $T$, we have found (by fitting the curves such as shown in Fig. 3a) that this increase is related to changes in $\mu_L$ rather than $\rho_S$.

Finally, ethanol has also offered the opportunity to significantly change $\kappa$ in situ by varying $T$. Fig. 3b shows $\mu_H$ for two devices immersed in ethanol as a function of $T$. As $T$ decreases, $\kappa$ increases reaching $\approx 55$ near the freezing point at 160 K. This increase in $\kappa$ was not accompanied by any significant changes in $\mu$ for all $n \geq 1\times10^{12}$ cm$^{-2}$ assessable through the Hall measurements, which again disagrees with the Coulomb scattering mechanism (see the theory curve in Fig. 3b). Furthermore, we studied several devices immersed in water and, also, covered with a liquid crystal MLC6204 ($\kappa \approx 44$). In the latter case, we used the back gate as for glycerol. The experiments in water required thorough electrical isolation and proved to be more difficult due to evaporation, when the back gate was used, and due to electrical erosion for the liquid gating. Only marginal changes in $\mu$ were observed for these dielectrics, too. This shows that Coulomb scatterers can certainly influence $\mu$ but do not limit it. Indeed, if charged impurities were limiting scatterers, $\mu$ should have increased by a factor of 5 in ethanol, 10 in MLC6204 and 20 in water [4,15].

To summarize, no one doubts that charged impurities are present in graphene devices. They are certainly responsible for the observed chemical shift of the NP and, at least partially, for electron-hole puddles. Also, one can imagine that Coulomb scatterers are dominant in some devices (especially, those exhibiting anomalously low $\mu$ ~100 to 1,000 cm$^2$/Vs as reported by several groups). In this case, high-$\kappa$ media are expected to increase their $\mu$ by an order of magnitude and make the resistance peak much sharper, similar to the behavior shown in Fig. 2a but with more pronounced changes. On the other hand, our results show that Coulomb scatterers are not the impurities that limits $\mu$ in typical devices with $\mu$ ~10,000 cm$^2$/Vs [1,2,5-7,10,11].

If not charged impurities, what could then be the limiting scatterers? A certain type of ripples (quenched flexural phonons) [17] can lead to long-range scattering, which is unaffected by dielectric environment. However, there is mounting experimental evidence that such ripples are probably not the dominant source of scattering. Another possibility is resonant scatterers with the energy close to the Dirac point. Such impurities can be common in graphene [18] and lead to $\rho \propto 1/n\ln^2(k_F R)$ which for an atomic scale potential of size $R$ results in a dependence that mimics $\sigma \propto n$ [19]. One can expect a smaller effect of high $\kappa$ media on resonant scatterers but there is no quantitative theory to compare with. If resonant scatterers attach strongly to graphene, their removal may require high-$T$ annealing by electric current (>1000 ºC), which so far was possible only for suspended samples [7]. In general, the scattering mechanism responsible for limited $\mu$ is open for discussion but the observed indifference to high $\kappa$ puts severe constraints on possible candidates.

Acknowledgements: We thank F. Guinea and S. Das Sarma for useful discussions. This work was supported by EPSRC (UK) and the Royal Society.

1. For review, see A. K. Geim, K. S. Novoselov, *Nature Mater.* **6**, 183 (2007); A. H. Castro Neto *et al*, *Rev. Mod. Phys.* **81**, 109 (2009).
2. K. S. Novoselov *et al*, *Nature* **438**, 197 (2005); Y. Zhang *et al*, *Nature* **438**, 201 (2005).
3. T. Ando, *J. Phys. Soc. Japan* **75**, 074716 (2006); K. Nomura, A. H. MacDonald, *Phys. Rev. Lett.* **96**, 256602 (2006).
4. S. Adam, E. W. Hwang, V. M. Galitski, S. Das Sarma, *Proc. Natl. Acad. Sci. U.S.A.* **104**, 18392 (2007).
5. F. Schedin *et al*, *Nature Mater.* **6**, 652 (2007).
6. J. H. Chen *et al*, *Nature Phys* **4**, 377 (2008).
7. K. Bolotin *et al*, *Solid State Commun.* **146**, 351 (2008); X. Du, I. Skachko, A. Barker, E. Y. Andrei, *Nature Nanotech.* **3**, 491 (2008).
8. J. Martin *et al*, *Nature Phys.* **4**, 144 (2008).
9. P. Blake *et al*, *Appl. Phys. Lett.* **91**, 063124 (2007).
10. S. V. Morozov *et al*, *Phys. Rev. Lett.* **100**, 016602 (2008).
11. Y. W. Tan *et al*, *Phys. Rev. Lett.* **99**, 246803 (2007).
12. BSCCO is a high $\kappa$ material, and the fact that no increase in $\mu$ is seen in Fig. 1 is indicative. However, we avoid using this as an argument because $\kappa$ for BSCCO is not well known [Z. Zhai *et al*, *Phys. Rev. B* **63**, 092508 (2001)].
13. D. Jena, A. Konar, *Phys. Rev. Lett.* **98**, 136805 (2007).
14. Our devices were annealed in hydrogen at ~350 Cº to make sure that no polymer film is left on the surface. AFM measurements showed graphene being <1 nm above the SiO$_2$ surface, which is typical for clean graphene. This is important because a few nm of a PMMA residue can effectively [15] decouple graphene from a dielectric deposited on top. Also, we have found that the top dielectric can induce significant changes in the back-gate capacitance. This was taken into account by measuring $\rho_{xy}$ which yields $n$ in Fig. 2a.

15. To estimate the effect of a dielectric film placed on top of graphene, we have considered the standard scattering problem for a Coulomb potential $V(q) \propto 1/q\varepsilon$. Here, $\varepsilon$ is the effective dielectric function that, because of a finite thickness $D$ of the film, is dependent on the in-plane electron wavevector $q$ [13] and is given by

$$\varepsilon(q) = \frac{F(q)}{2G(q)} \text{ with } F(q) = \kappa + \kappa_S - \left[\frac{(\kappa-1)(\kappa-\kappa_S)}{(\kappa+1)}\right]\exp(-2qD) \text{ and } G(q) = 1 + \left[\frac{\kappa-1}{\kappa+1}\right]\exp(-2qD).$$

This formula is a straightforward solution for the three media electrostatic problem with a charge placed at the interface between the SiO$_2$ substrate (dielectric constant $\kappa_S$) and the film. One can check that this yields $\varepsilon = (\kappa_S + 1)/2$ and $\varepsilon = (\kappa_S + \kappa)/2$ for $D = 0$ and $\infty$, respectively. The scattering potential $V(q)$ is then used within the Born approximation [3,4], which leads to the expression

$$S \propto \int_0^1 \frac{dx\, x^2 \sqrt{1-x^2}}{\left[x\varepsilon(2k_F x) + \frac{2e^2}{\hbar v_F}\right]^2}$$

where the suppression factor $S$ describes changes in $\rho$ as a function of $D$, $\kappa$ and the Fermi wavevector $k_F$. For ethanol, glycerol and water on top of graphene at room $T$, $S$ yields an increase in $\mu$ by ~5, 10 and 22 times, respectively. These values are in agreement with ref. [4]. Note that it requires a few nm of dielectric on top of graphene to approach the limit $D = \infty$. Therefore, it is incorrect to consider 1 to 2 atomic layers of ice as a semi-infinite medium as done by C. Jang *et al*, *Phys. Rev. Lett.* **101**, 146805 (2008) who observed a small (30%) increase in $\mu$ due to such an ice layer and interpreted this as evidence for charged impurities.

16. M. Krüger, M. R. Buitelaar, T. Nussbaumer, C. Schoenenberger, L. Forro, *Appl. Phys. Lett.* **78**, 1291 (2001).
17. M. I. Katsnelson, A. K. Geim, *Phil. Trans. R. Soc. A* **366**, 195 (2008).
18. T. O. Wehling *et al*, *Nano Lett.* **8**, 173 (2008); *Phys. Rev. B* **75**, 125425 (2007).
19. M. I. Katsnelson, K. S. Novoselov, *Solid State Commun.* **143**, 3 (2007); T. Stauber, N. M. R. Peres, F. Guinea, *Phys. Rev. B* **76**, 205423 (2007); P. M. Ostrovsky, I. V. Gornyi, A. D. Mirlin, *Phys. Rev. B* **74**, 235443 (2006).